\documentclass[letter,longauth]{aa} 
\usepackage{graphicx}
\usepackage{amssymb}
\newcommand{\hi}{{\rm H\,}{{\sc i}}}
\usepackage{txfonts}
\begin{document}
   \title{Cold dust clumps in dynamically hot gas\thanks{Herschel is an ESA space observatory with science instruments provided by
   European-led Principal Investigator consortia and with important
   participation from NASA.}
   }

   \author{S. Kim\inst{1}
    \and     E. Kwon\inst{1}
    \and     S.C. Madden\inst{2}
    \and     M. Meixner\inst{3,19}
    \and     S. Hony\inst{2}
    \and     P. Panuzzo\inst{2}
    \and     M. Sauvage\inst{2}
    \and     J. Roman-Duval\inst{3}
    \and     K.D. Gordon\inst{3}
    \and     C. Engelbracht\inst{4}
    \and     F.P. Israel\inst{5}
    \and     K. Misselt \inst{4}
    \and     K. Okumura\inst{2}
    \and     A. Li\inst{6}
    \and     A. Bolatto\inst{7}
    \and     R. Skibba\inst{4}
    \and     F. Galliano\inst{2}
    \and     M. Matsuura\inst{8,9}
    \and     J.-P. Bernard\inst{10}
    \and     C. Bot\inst{11}
    \and     M. Galametz\inst{2}
    \and     A. Hughes\inst{12,13}
    \and     A. Kawamura\inst{14}
    \and     T. Onishi\inst{15}
    \and     D. Paradis\inst{16}
    \and     A. Poglitsch\inst{17}
    \and     W. T. Reach\inst{16,18}
    \and     T. Robitaille\inst{19}
    \and     M. Rubio\inst{20}
     \and    A.G.G.M. Tielens\inst{5}
}
\institute{ Astronomy \& Space Science, Sejong University, 143-747, Seoul, South Korea    \\
\email{sek@sejong.ac.kr}
\and CEA, Laboratoire AIM, Irfu/SAp, Orme des Merisiers, F-91191 Gif-sur-Yvette, France
\and Space Telescope Science Institute, 3700 San Martin Drive, Baltimore, MD 21218, USA
\and Steward Observatory, University of Arizona, 933 North Cherry Ave., Tucson, AZ 85721, USA
\and Sterrewacht Leiden, Leiden University, P.O. Box 9513, NL-2300 RA Leiden, The Netherlands
\and 314 Physics Building, Department of Physics and Astronomy, University of Missouri-Columbia, Columbia, MO 65211, USA
\and Department of Astronomy, Lab for Millimeter-wave Astronomy, University of Maryland, College Park, MD 20742-2421, USA
\and Department of Physics and Astronomy, University College London, Gower Street, London WC1E 6BT, UK
\and Mullard Space Science Laboratory, University College London, Holmbury St. Mary, Dorking, Surrey RH5 6NT, United Kingdom
\and Centre d' \'{E}tude Spatiale des Rayonnements, CNRS, 9 av. du Colonel Roche, BP 4346, 31028 Toulouse, France
\and Observatoire Astronomique de Strasbourg, 11, rue de 1'universite, 67000 STRASBOURG, France
\and Center for Supercomputing and Astrophysics, Swinburne University of Technology, Hawthorn VIC 3122, Australia
\and CSIRO Australia Telescope National Facility, 76 Epping Rd.,  NSW1710, Australia
\and Department of Astrophysics, Nagoya University, Chikusa-ku, Nagoya 464-8602, Japan
\and Department of Physical Science, Osaka Prefecture University, Gakuen 1-1, Sakai, Osaka 599-8531, Japan
\and Spitzer Science Center, California Institute of Technology, MS 220-6, Pasadena, CA  91125, USA
\and Max-Planck-Institut f$\mathrm{\ddot{u}}$r extraterrestrische Physik, Giessenbachstra 85748 Garching,  Germany
\and Stratospheric Observatory for Infrared Astronomy, Universities Space Research Association, Mail Stop 211-3, Moffett Field, CA 94035, USA
\and   Center for Astrophysics, 60 Garden St., MS 67 , Harvard University, Cambridge, MA 02138, USA
\and Departamento de Astronomia, Universidad de Chile, Casilla 36-D, Santiago, Chile
}

 \abstract
 {}
{We present clumps of dust emission from Herschel observations of the Large Magellanic Cloud (LMC) and their
physical and statistical properties.
We catalog cloud features seen in the dust emission from Herschel observations of the LMC, the Magellanic type irregular galaxy closest to the Milky
Way, and compare these features with \hi\ catalogs from the ATCA+Parkes \hi\ survey.}
{Using an automated cloud-finding algorithm, we identify clouds and clumps of dust emission and examine the cumulative mass distribution of the detected dust clouds.
The mass of cold dust is determined from physical parameters that we derive by performing spectral energy distribution fits to 250, 350, and 500 $\mu$m emission from SPIRE observations using dust grain size distributions for graphite/silicate in low-metallicity extragalactic environments.}
{The dust cloud mass spectrum follows a power law distribution with an exponent of
$\gamma=-1.8$ for clumps larger than 4$\times$10$^2$ M$_\odot$ and is similar to the
\hi\ mass distribution. This is expected from the theory of ISM structure in the vicinity of star formation.}
{}
                \authorrunning
                \titlerunning
   \maketitle

\section{Introduction}

The overall structure and evolution of the ISM of a galaxy is determined by the physical
and dynamical conditions in the diffuse ISM where most of the processing of gas and dust
clouds occurs. Thus inspecting the cloud properties by studying the atomic gas and the dust emission under different environmental conditions will help us to understand how this processing of matter proceeds.

In the present paper, we study the dynamical conditions of the diffuse
gas in the Large Magellanic Cloud (LMC). We examine optically thin dust emission at far-infrared (FIR) and sub-millimeter wavelengths by studying the spectral energy
distributions (SEDs) inferred from radiative transfer models.
By combining an \hi\ synthesis survey and an imaging survey in the FIR and submillimeter emission performed by {\it Herschel} Space Observatory (Pilbratt et al. 2010), we investigate the
detailed relationship between the diffuse atomic gas and interstellar dust emission in the
LMC. The {\it Herschel} Space Observatory is designed to study the submillimeter and FIR emission from our universe and offers unprecedented resolution. The {\it Herschel} has a payload of
two cameras, PACS (Poglitsch et al. 2010) and SPIRE (Griffin et al. 2010). PACS employs two bolometer arrays for imaging, and SPIRE comprises spider-web bolometers with NTD Ge temperature
sensors. Mapping strategy and data processing procedures are described in detail by Meixner et al. (2010).

The {\it Herschel} survey of the Magellanic Clouds will provide information about the interstellar dust emission (Meixner et al. 2006) from all phases of the ISM in the
Magellanic Clouds. For diffuse atomic gas, we use data from the \hi\ aperture synthesis survey in Kim et al. (1998) and data from the combined survey using single-dish observations with the Parkes 64-m radio telescope (Staveley-Smith et al. 2003) that was published in Kim et al. (2003).
The SED models provide spectral information in the 160 $\mu$m to 500 $\mu$m
wavelength range. Relatively optically thin dust emission at
FIR and submillimeter wavelengths can also provide information about the gas mass in the ISM because the gas and dust are well mixed in most of the ISM phases. Therefore, the dust emission at the FIR and submillimeter wavelengths can trace the column density and the structure of the ISM.

\begin{figure}
\scalebox{0.45}
   {\includegraphics{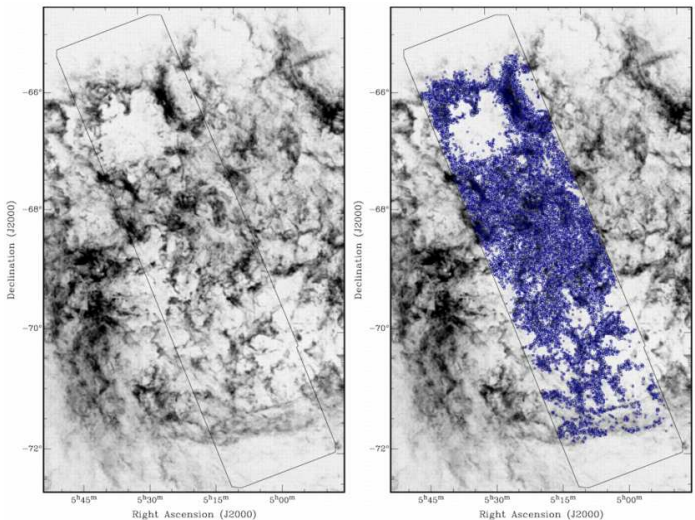}}
 \caption[width=12cm]{Dust clouds seen in the SPIRE 500 $\mu$m image were identified
   using an automatic finding algorithm in which the lowest threshold was set to be 3 $\sigma$ (blue contours), and they are overlaid on the \hi\ aperture synthesis image (Kim et al. 2003) of the LMC. 1$\sigma$ is set to 0.3 MJy/sr at 500 $\mu$m. The black box indicates the
   observed region in the Science Demonstration Program of Herschel Space Observatory
   (Meixner et al. 2010).\\}
 \end{figure}

\begin{figure}
\scalebox{0.46}
{\includegraphics{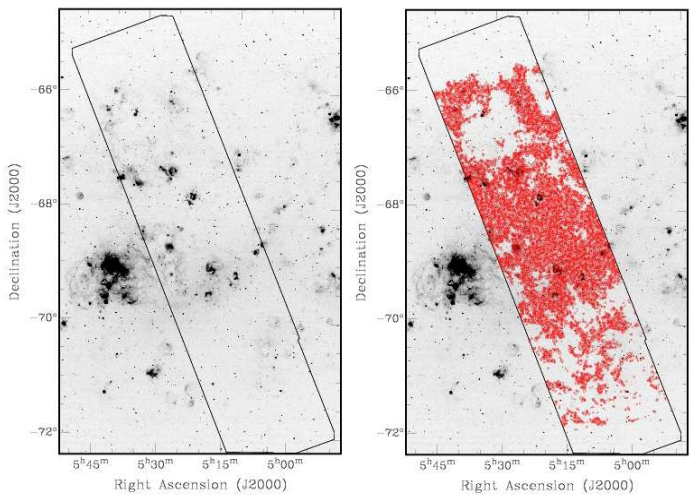}}
   \caption[width=12cm]{Dust clouds seen in the SPIRE 350 $\mu$m image were identified using an
   automatic finding algorithm, the lowest threshold being set to 3 $\sigma$ (red contours), and
   overlaid on the H$\alpha$ image of the LMC (Kim et al. 1999). The black box indicates the
   observed region in the Science Demonstration Program of Herschel Space Observatory (Meixner
   et al. 2010).\\}
   \end{figure}

\section{Cold dust associated with atomic hydrogen}
\subsection{Dust clumps}

The unprecedented quality of the SPIRE images improves our knowledge of the spatial
distribution of galactic infrared/submillimeter dust radiation and characterizes it as
cloud features with clumps and filaments. We adopt the 'clump' terminology to represent
entities with properties of typical 'clouds' seen in the molecular clouds in the Galaxy
(Bergin and Tafalla 2007). We examine the brightness distribution of the pixels treating them as clumps
by searching for peaks of local emission and adding the neighboring pixels of one clump to the objects (Williams, de Geus, \& Blitz 1994). We find 7449 dust clumps generated by
an automatic clump-finding routine for the 500 $\mu$m emission. The threshold being set to
three times of an RMS of $\sim$0.3 MJy/sr (Fig. 1). The automatic clump-finding algorithm determines structure by first contouring the data at a multiple of the RMS noise of the data, searching for peaks of emission that are local maxima in the SPIRE images, and then connecting pixels at each contour level from the highest to lower intensities. Isolated contours at each level are identified as clumps (Williams, de Geus, \& Blitz 1994).

By applying the same method to the 350 $\mu$m image of the LMC, we find 8460 dust
clumps (Fig. 2). More faint clumps are seen at 350 $\mu$m than 500 $\mu$m above
the noise level. However, the dust clumps at 500 $\mu$m are well correlated with
the \hi\ clouds and filaments. In general, the dust clumps at 500 $\mu$m identified
 in the present study are
distributed quite uniformly in the LMC. Thus, for the clump analysis pursued in this study
we use the clumps identified in the 500 $\mu$m image.

For a clump consisting of a set of pixels with positions $\{x,y\}$ and intensity $\{I\}$,
the size of each clump is calculated based on its extent in the spatial dimension below.
The clumps range in size from 9.8 to 47$\pm$1 pc with a median of 15$\pm$1 pc in radius given by

\begin{equation}
\sigma_{x}^{2} = \left[\frac{\sum{Ix^2}}{\sum{I}}-\left(\frac{\sum{Ix}}{\sum{I}}\right)^2\right].
\end{equation}
\noindent

\subsection{Physical properties of dust clumps}
Each dust clump was characterized by the radiation source properties associated with the
dust clump and their ambient medium using DUSTY radiative transfer calculations performed by Ivezi${\rm\acute{c}}$ et al. (1999). This model assumes a spherical geometry and supports
an analytical form of the dust density distribution (Sarkar et al. 2006). The physical
parameters constrained by fitting the observed SEDs include the chemical composition of the dust, grain size distribution, dust temperature at the
inner and outer boundary, dust density distribution, optical depth at a specific wavelength,
 and the ambient interstellar radiation field (ISRF). We adopted the chemical properties of the dust determined by Draine \& Li (2001) and Weingartner \& Draine (2001). For dust in the neutral medium, we
 adopted the size distribution of Kim, Martin, \& Hendry (1994, KMH) using the power law
 distribution function for grain sizes, $n(a) \propto a^{-q}e^{-a/a_0}$ with $q$=3.5,
 $a_{min}$=0.005 $\mu$m, and $a_{max}$=0.25 $\mu$m. The classical model of grain size
 distribution was constructed by Mathis, Rumpl, \& Nordsieck (1977, MRN), and Draine and
 Lee (1984) revised the MRN model with dielectric functions for graphite and silicate (Draine
 and Li 2001).

Interstellar grain temperatures are calculated from 
\begin{equation}
\int_{0}^{\infty}\pi B_\lambda(T_d)Q_{IR}(a,\lambda)d\lambda = \int_{912\AA}^{\infty}F_{\lambda}Q_{UV}(a,\lambda)d\lambda
\end{equation}
where $B_\lambda(T)$ is the Planck function for dust at temperature $T$, $F_\lambda$ is
the flux of the interstellar radiation field,
and $Q_{IR}(a,\lambda)$ and $Q_{UV}(a,\lambda)$
indicate the infrared and ultraviolet absorption efficiencies, respectively, and are
functions of the dust grain radius $a$ (Spitzer 1978). The ISRF is produced by three components: (1) emission from stars, (2) dust emission in the
far-IR associated with the heating of interstellar matter by absorption of star light, and
(3) the cosmic background radiation (Cox et al. 1986; Chi \& Wolfendale 1991; Strong et al.
2000). Unlike the Milky Way, confusion along the line-of-sight to the LMC is negligible because of its low inclination (Weinberg and Nikolaev 2001). The LMC also has low internal
reddening, and its proximity permits the detailed study of its stellar population. Therefore, the ISRF in the LMC can be estimated with confidence.

The dust equilibrium temperatures are obtained by determining the energy balance between absorption
and emission. In Fig. 3, we present three examples of applying DUSTY to different clumps. The result of each fitting provides a characteristic dust emission temperature, $T_d$ for each clump and an estimate of the graphite and silicate abundance ratio. The resulting graphite and silicate abundances from fitting the observed SEDs differ by 50\% among the clumps in
different environments in the LMC. The second approach to characterize clump temperature is to apply GRASIL (Silva et al. 1998). This spectrophotometric self-consistent model
computes the absorption and emission by dust in three different environments such as
molecular clouds, diffuse ISM, and AGB envelopes with a dust model consisting of big
grains, small grains, and polycyclic aromatic hydrocarbons (PAH) molecules. The consistency check on the equilibrium temperature is made using the GRASIL model and the observed SEDs (Fig. 3). In general, the observed fluxes at 160 $-$ 500 $\mu$m are ascribed to the dust in the diffuse medium heated by the interstellar radiation field.
Heating of diffuse gas in the ISM can occur by means of the photoelectric heating in interstellar clouds where the dust grains act as catalytic surfaces. The resultant temperature from the fit to the observed SEDs ranges from 15$\pm$0.4 to 25$\pm$0.4 K and is similar to the SPIRE temperature map (Gordon et al. 2010).

At a given grain temperature, the mass of each dust clump is calculated from
$M_{d} = {S_{\lambda}{\times} D^{2}}/{{\kappa}_{\lambda}{\times} B_{\lambda}(T_d)}$,
where $S_{\lambda}$ is the flux density at 500 $\mu$m and $D$ is the
distance of the galaxy at 50 kpc (Schaefer 2008), and $B_{\lambda}(T_d)$ is the value of the Planck function at 500 $\mu$m, and a function of $T_d$. We adopt a mean temperature for each clump as a function of the three categories shown in Fig. 3. The YSOs region and bar region in Fig. 3 indicate
dynamically hot regions described in Kim et al. (1998, 2007),
where $\kappa_{\lambda}$ is the mass absorption coefficient at 500 $\mu$m from the SPIRE observations and $\kappa_{\lambda}\sim1.15$ cm$^2$/g is used. The derived dust masses are in the range of 1.8 $\times$ 10$^1$ M$_\odot$ $<$ M$_{dust}$ $<$ 7.9 $\times$10$^3$ M$_\odot$.

\begin{figure}
\vspace{0.5cm}
\includegraphics[width=8cm]{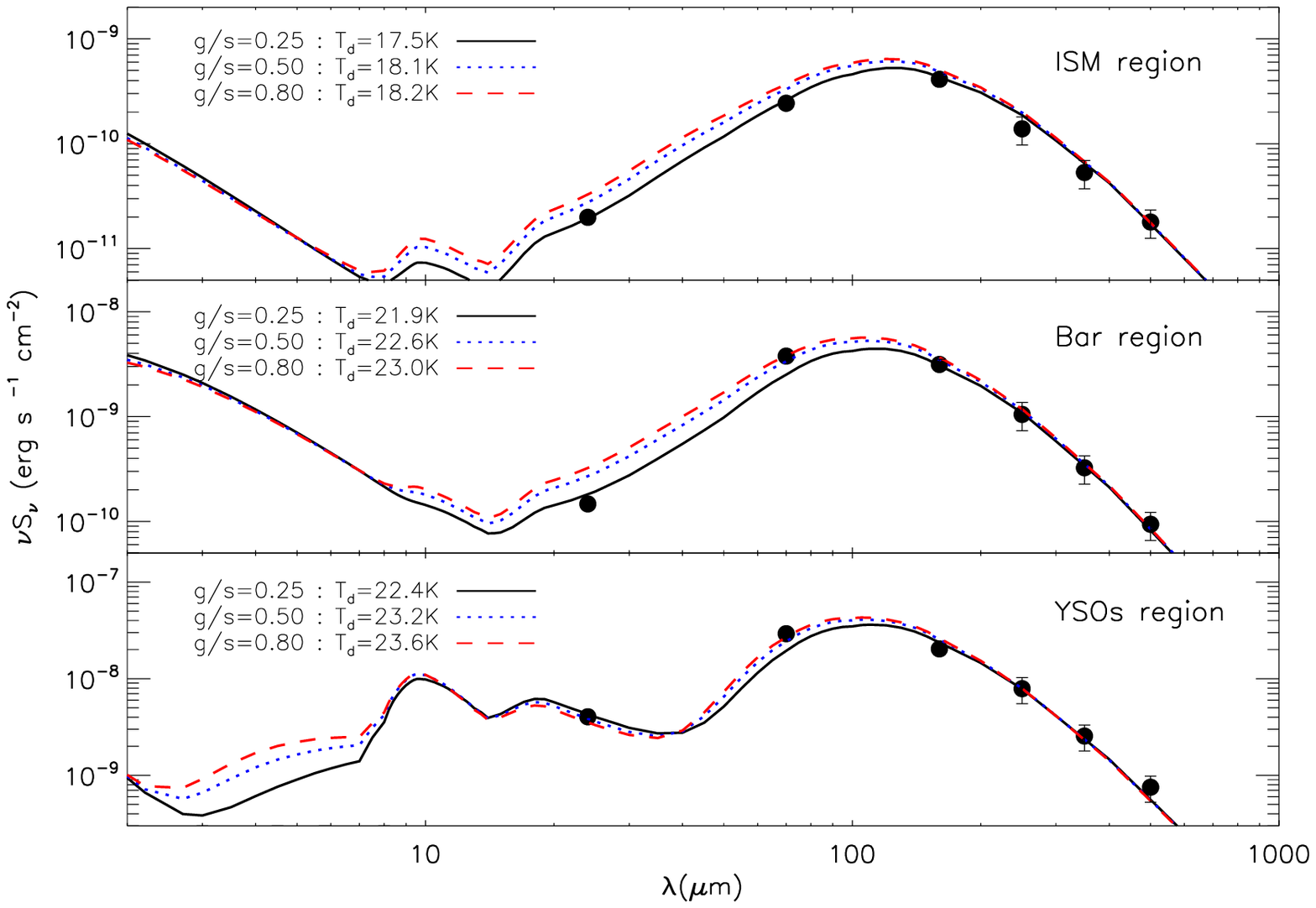}
\includegraphics[width=8cm]{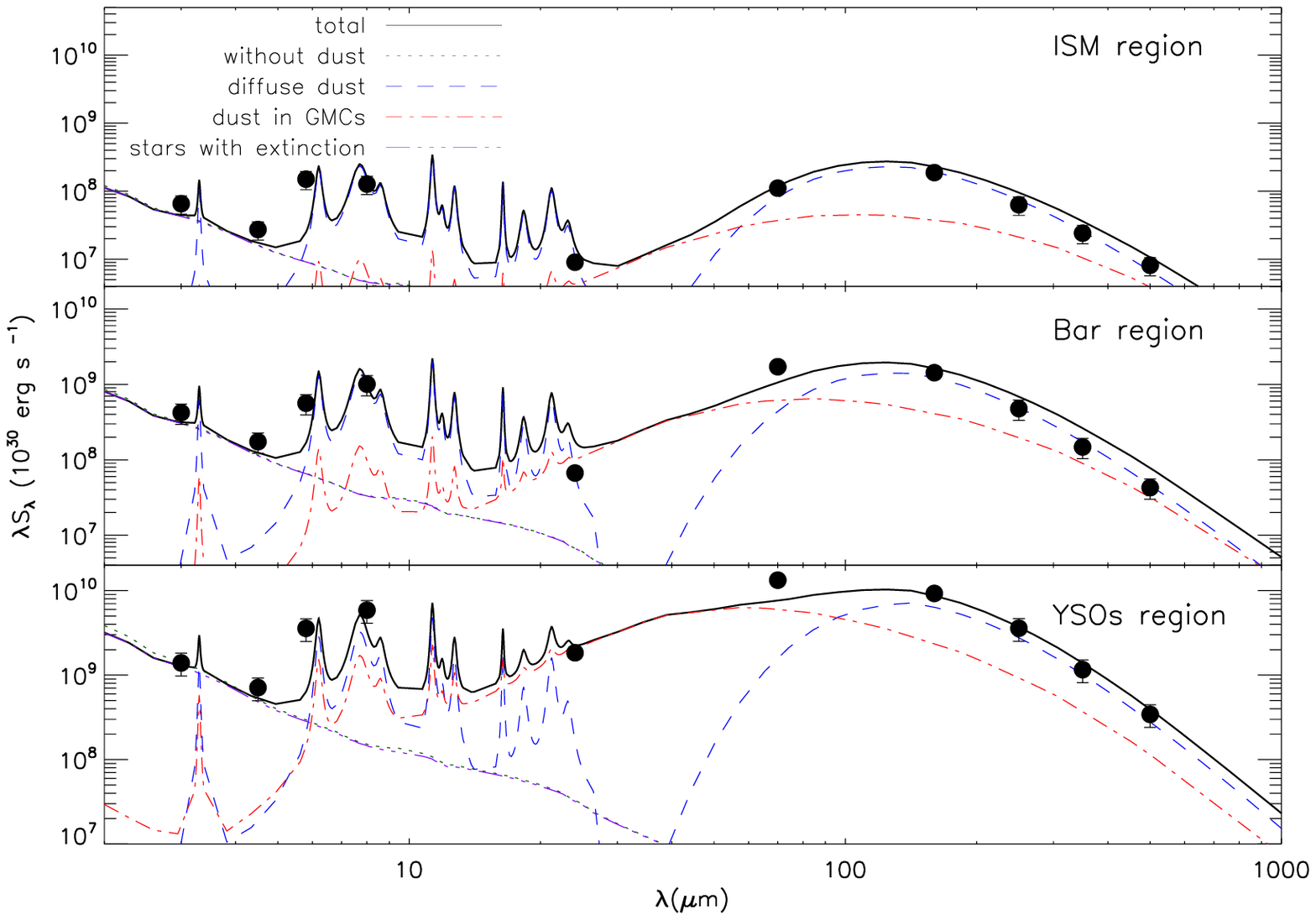}
\caption[width=12cm]{DUSTY model spectra and the best SED fitting results for three
representative dust clumps in different environments (Top). The ISRF for each clump
is 1.0 $G_0$ (ISM region), 3.5 $G_0$ (Bar region), and 7.0 $G_0$ (YSOs region). GRASIL
(Silva et al. 1998) model spectra and the observed SEDs (Bottom). The data points are
from IRAC, MIPS, and SPIRE.}
\end{figure}

To characterize the clump properties, the mass spectrum of the clumps is examined.
We choose the cumulative spectrum rather than the differential mass distribution since
the relative masses will be more or less the same, when the mass range is small over
the entire spectrum. For each catalog, we fit a power law function to the cumulative
mass distribution (Rosolowsky et al. 2005; Kim et al. 2007):

\begin{equation}
N(M'>M)=N_0\bigg(\frac{M}{10 M\odot}\bigg)^{\gamma +1},
\end{equation}
where $N_0$ is the number of clouds in the derived distribution with masses higher than $10$ $M_\odot$ and $\gamma$ is the index of the differential mass distribution. The
cumulative mass distribution of dust cores in the catalog of sources identified in
this study and the residuals of the fit are presented
in Fig. 4. The clump mass spectrum follows a broken power law in the mass distribution. For masses
lower than 4 $\times$ 10$^2$ M$_\odot$, it follows a power law distribution and decreases
slowly with an exponent of the mass distribution, $\gamma=-0.8$. For the clumps more massive than 4 $\times 10^2$ M$_\odot$, the mass spectrum is significantly steeper with a slope
of $\gamma=-1.8\pm0.1$ than in the lower mass regime. The observed clump mass spectrum for 1.8 $\times$ 10$^1$ M$_\odot$ $<$ M$_{dust}$ $<$ 4.0 $\times$10$^2$ M$_\odot$, exhibits
a flatter power law than the stellar initial mass function (Salpeter 1955). This confirms
that most of the clump mass resides in massive clumps, while the most of the stellar mass
is in low-mass stars (McKee and Williams 1997; Johnstone et al. 2000).

\begin{figure}
\vspace{0.5cm}
\scalebox{0.55}
{\includegraphics[65,100][500,320]{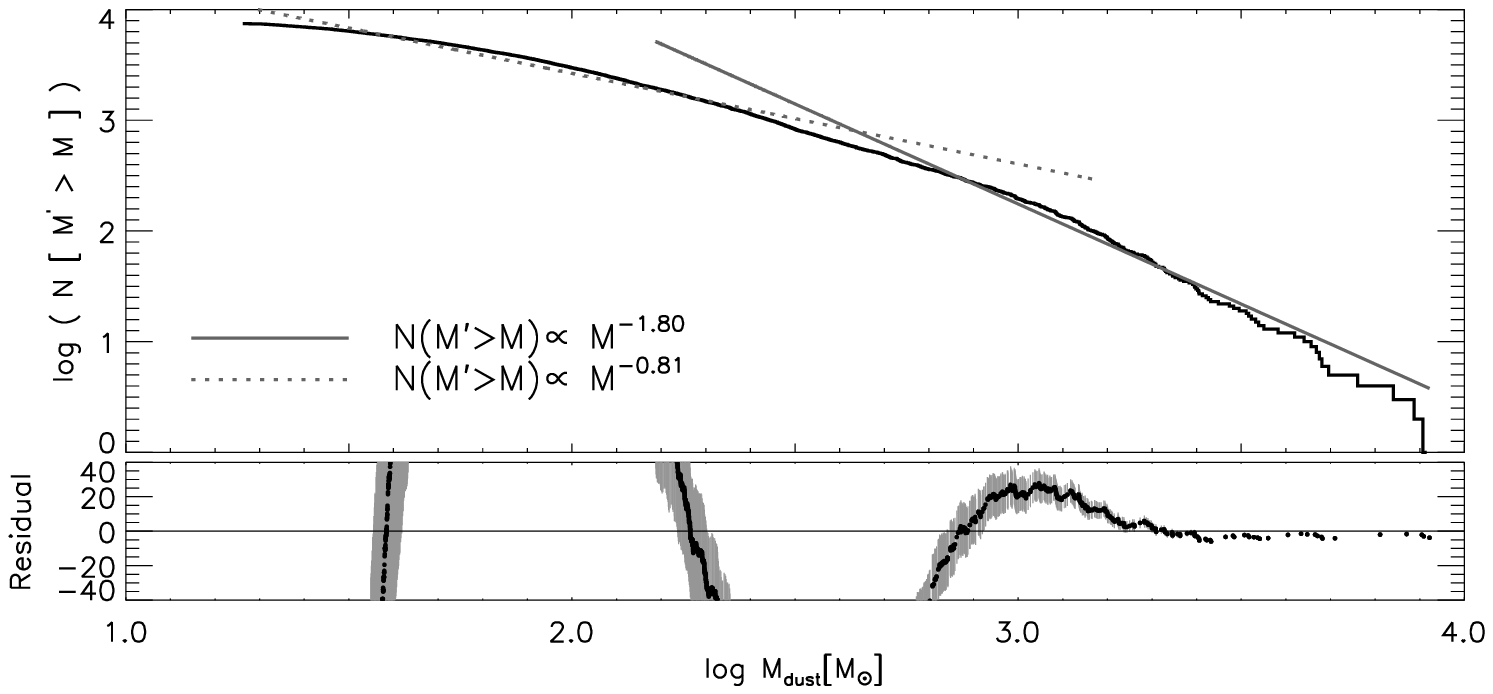}}
\scalebox{0.55}
{\includegraphics[65,88][500,320]{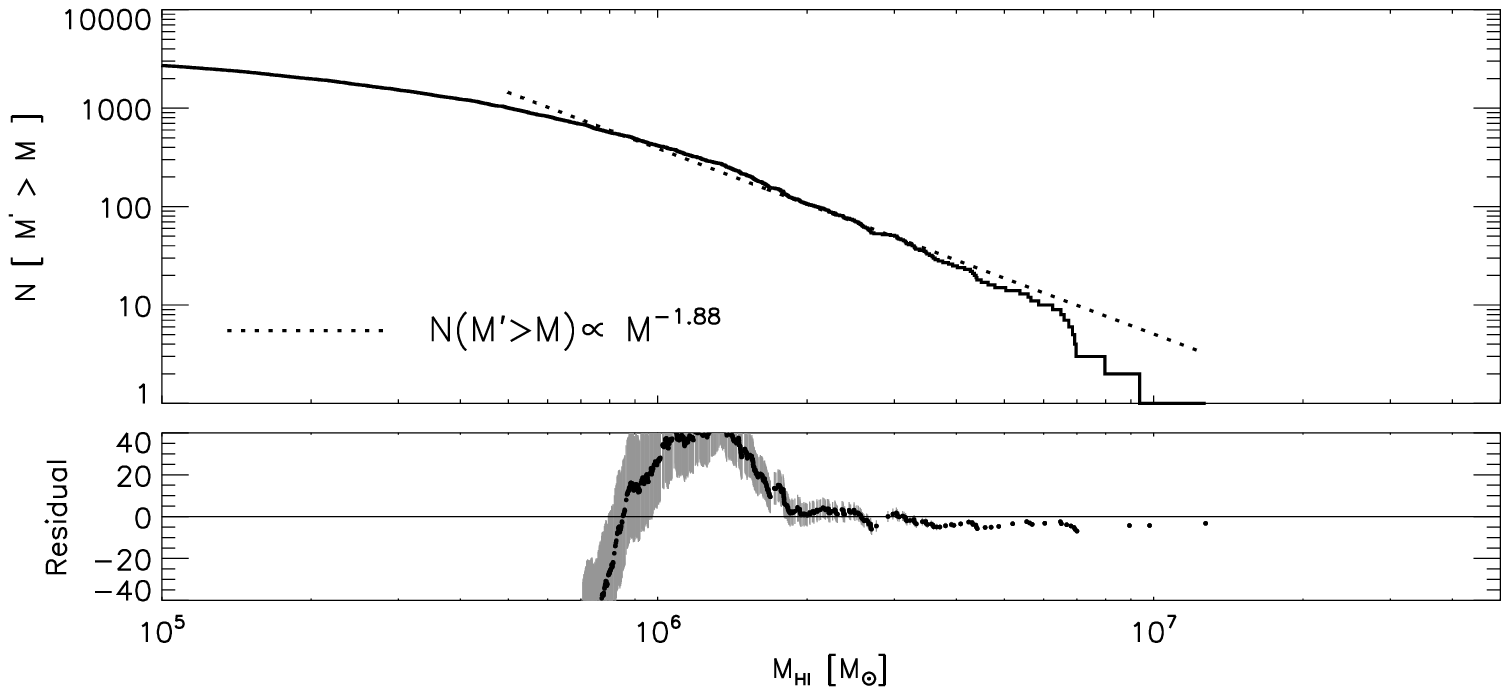}}
\caption[width=12cm]{Dust clump mass spectrum (Top). The slope of the mass spectrum,
$\gamma$, is -1.8 with an error estimate of $\pm$0.1. \hi\ clump mass spectrum (Bottom).
The slope of the \hi\ mass spectrum, $\gamma$, is -1.88 with an error estimate of
$\pm$0.1 for the fit in the selected mass range.}
\end{figure}

\subsection{Gas and dust mass spectrum}

Observation and analysis of the dust clumps and neutral cloud structure indicate that a
number of smaller clumps reside in the neutral diffuse clouds. The physical properties
of dust clumps were characterized in terms of their mass spectrum, which follows a power law distribution. We note that the distribution of dust clump mass is quite similar to the
\hi\ mass distribution, the index of the mass distribution being, $\gamma=-1.88$. The mass of the \hi\ cloud was determined from its integrated intensity $I$ measured in K~km~s$^{-1}$

\begin{equation}
M_{cloud} = 1.822\times 10^{18} I \sigma_{cl} m_{\mathrm{H}}, 
\end{equation}

where $\sigma_{cl}$ is the total cloud area in cm$^2$.
Since the dust emission in the far-IR continuum emission provides an estimate of the
star formation rate (SFR) and grains remain strongly coupled to the neutral gas, we expect the mass spectra of the dust clouds to behave in a similar way to the mass spectra
resulting in the diffuse gas.

The mass function near the high-end appears to be even steeper with
$\gamma=-2.1$. The index of that power law fit can be steeper where star formation is ongoing, such that cloud dissipation occurs (Wada et al. 2000). The similarity of clump mass
distribution suggests that the overall gas-to-dust mass ratio is more or less uniform
in the region considered here. This is consistent with the overall gas-to-dust mass ratio being relatively uniform across the LMC, except for the super-shell affected by the
supernova shocks (Gordon et al. 2003). A detailed analysis of
the gas-to-dust mass ratio of two molecular clouds in the LMC can be found in Roman-Duval et al. (2010).

\section{Summary}

Dust clumps have been identified and cataloged in the {\it Herschel} SPIRE survey of the LMC using an automated cloud-finding
algorithm. The distribution of cold dust clumps is remarkably similar to the \hi\ clump mass distribution, sharing an index of mass distribution, $\gamma=-1.8$. However, the dust clump mass spectrum in the lower mass regime follows a flatter power law than the Salpeter stellar IMF.

\begin{acknowledgements}
    We acknowledge financial support from the NASA Herschel Science Center, JPL contracts
    \# 1381522 \& 1381650.  We thank the contributions and support from the European Space
    Agency (ESA),  the PACS and SPIRE teams, the Herschel Science Center and the NASA
    Herschel Science Center (esp. A. Barbar and K. Xu) and the PACS and SPIRE instrument
    control centers,  without which none of this work would be possible. We thank the referee for his/her very important comments on the manuscript. SK and EK were
    supported by Basic Science Research Program through the National Research Foundation
    of Korea (NRF) funded by the Ministry of Education, Science and Technology 2009-0062866.
    M.R. is supported by FONDECYT No1080335 and FONDAP No15010003.

\end{acknowledgements}


\begin{thebibliography}{}

\bibitem[2007]{Bergin}Bergin, E. A., \& Tafalla, M. 2007, ARA\&A, 45, 339B

\bibitem[1991]{Chi}Chi, X., Wolfendale, A. W. 1991, ICRC, 2, 233C

\bibitem[1986]{Cox}Cox, P., Kruegel, E., \& Mezger, P. G. 1986, A\&A, 155, 380C

\bibitem[1984]{Draine}Draine, B. T., \& Lee, H. M. 1984, ApJ, 285, 89D

\bibitem[2001]{Draine}Draine, B. T., \& Li, A. 2001, ApJ, 551, 807D

\bibitem[2003]{Gordon}Gordon, K. D., Clayton, G.C., Misselt, K.A., Landolt, A.U., \& Wolff, M.J. 2003, ApJ, 594, 279

\bibitem[2010]{Gordon}Gordon, K. D. et al. 2010, this volume

\bibitem[2010]{Griffin}Griffin, M. et al. 2010, this volume

\bibitem[1999]{Ivezic}Ivezic, Z., Nenkova, M., \& Elitzur, M. 1997, MNRAS, 287, 799

\bibitem[2000]{Johnstone} Johnstone, D., Wilson, C.D., Moriarty-Scieven, G., Joncas, G., Smith, G., Gregersen, E., \& Fich, M. 2000, ApJ, 545, 327J

\bibitem[1994]{Kim}Kim, S., Martin, P. G., \& Hendry, P.D. 1994, ApJ, 422, 164K

\bibitem[1998]{Kim}Kim, S., Staveley-Smith, L., Dopita, M.A., Freeman, K.C., Sault, R.J., Kesteven, M.J., \& McConnell, D. 1998, ApJ, 503, 674K

\bibitem[1999]{Kim}Kim, S., Dopita, M., Staveley-Smith, L., \& Bessell, M. 1999, AJ, 118,2797
\bibitem[2003]{Kim}Kim, S., Staveley-Smith, L., Dopita, M.A., Sault, R.J., Freeman, K.C. Lee, Y., \& Chu, Y.-H. 2003, ApJS, 148, 473K

\bibitem[2007]{Kim} Kim, S., Rosolowsky, E., Lee, Y., Kim, Y.H., Dopita, M.A.,
  et al. 2007, ApJS, 171, 419

\bibitem[1977]{Mathis}Mathis, J. S., Rumpl, W., Nordsieck, \& K. H. 1977, ApJ, 217, 425M

\bibitem[1997]{McKee} McKee, C. F., \& Williams, J. P. 1997, ApJ, 476, 144M

\bibitem[2006]{Meixner} Meixner, M., Gordon, K., \& Indebetouw, R. et al. 2006, AJ, 132, 2268

\bibitem[2010]{Meixner} Meixner, M. et al. 2010, this volume

\bibitem[2010]{Pilbratt} Pilbratt, G. et al. 2010, this volume

\bibitem[2010]{Poglitsch} Poglitsch, A. et al. 2010, this volume

\bibitem[2010]{Roman-Duval} Roman-Duval, J. et al. 2010, this volume

\bibitem[2005]{Rosolowsky} Rosolowsky, E. 2005, PASP, 117, 1403

\bibitem[1955]{Salpeter} Salpeter, E. E. 1955, ApJ, 121, 161

\bibitem[2006]{Sarkar} Sarkar, G., \& Sahai. R. 2006, ApJ, 644, 1171

\bibitem[1998]{silva} Silva, L., Granato, G., Bressan, A., \& Danese, L. 1998, ApJ, 509, 103

\bibitem[1978]{Spitzer} Spitzer, L. 1978, New York Wiley-Interscience, 1978.~333 p.

\bibitem[2003]{Stavele} Staveley-Smith, L., Kim, S., Calabretta, M., Haynes, R., \& Kesteven, M. 2003, MNRAS, 339, 87

\bibitem[2000]{Strong} Strong, A. W., Moskalenko, I. V., \& Reimer, O. 2000, ApJ, 573, 763

\bibitem[2000]{wada} Wada, K., Spaans, M., \& Kim, S. 2000, ApJ, 540, 797

\bibitem[2001]{weinberg} Weinberg, M.D., \& Nikolaev, S. 2001, ApJ, 548, 712

\bibitem[2001]{weingartner} Weingartner, J.C., \& Draine, B.T. 2001, ApJ, 548, 296

\bibitem[2000]{williams} Williams, J., de Geus, E.J., \& Blitz, L. 1994, ApJ, 428, 693

\end{thebibliography}
\end{document}